\documentclass[aps, prl, superscriptaddress, twocolumn, showpacs, longbibliography, footnoography]{revtex4-1}
\usepackage{graphicx}
\usepackage{amsmath,amssymb}
\usepackage{mathtools,mathrsfs,dsfont}
\usepackage{bm, physics}
\usepackage[dvipsnames]{xcolor}
\usepackage[colorlinks, linkcolor=Red, citecolor=NavyBlue, urlcolor=NavyBlue]{hyperref}
\usepackage[all]{hypcap} 

\begin{document}

\title{Non-Hermitian Multipole Skin Effects Challenge Localization}
\author{Jacopo Gliozzi}
\affiliation{Institute for Condensed Matter Theory, University of Illinois Urbana-Champaign, Urbana, USA}
\author{Federico Balducci}
\affiliation{Max Planck Institute for the Physics of Complex Systems, Dresden, Germany}
\author{Taylor L. Hughes}
\affiliation{Institute for Condensed Matter Theory, University of Illinois Urbana-Champaign, Urbana, USA}
\author{Giuseppe De Tomasi}
\affiliation{CeFEMA-LaPMET, Departamento de F\'isica, Instituto Superior T\'ecnico, Universidade de Lisboa, Portugal}
\begin{abstract}
We study the effect of quenched disorder on the non-Hermitian skin effect in systems that conserve a U(1) charge and its associated multipole moments.
In particular, we generalize the Hatano-Nelson argument for a localization transition in disordered, non-reciprocal systems to the interacting case. 
When only U(1) charge is conserved, we show that there is a transition between a skin effect phase, in which charges cluster at a boundary, and a many-body localized phase, in which charges localize at random positions. 
In the dynamics of entanglement, this coincides with an area to volume law transition.
For systems without boundaries, the skin effect becomes a delocalized phase with a unidirectional current. 
If dipoles or higher multipoles are conserved, we show that the non-Hermitian skin effect remains stable to arbitrary disorder. 
Counterintuitively, the system is therefore always delocalized under periodic boundary conditions, regardless of disorder strength. 
\end{abstract}

\maketitle
\textit{Introduction}.---The theory of Anderson localization is a milestone in condensed matter physics that describes how a disordered potential can trap non-interacting quantum particles~\cite{Anderson1958}. Its generalization to interacting systems---dubbed many-body localization (MBL)---challenges our understanding of how the laws of statistical mechanics emerge in a closed system~\cite{BASKO_2006,Gornyi_2005,Nandkishore_2015,Abanin_2019,Sierant_2025}. In practice, however, complete isolation is only approximate. Interactions between a quantum system and its environment can induce dissipation, decoherence, and wavefunction collapse, destroying unitarity. Many of the resulting effects can be captured by non-Hermitian Hamiltonians~\cite{Zhang2022review,Ashida2020NonHermitian}.

\begin{figure}[t!]
\label{fig:1}
\includegraphics[width=\linewidth]{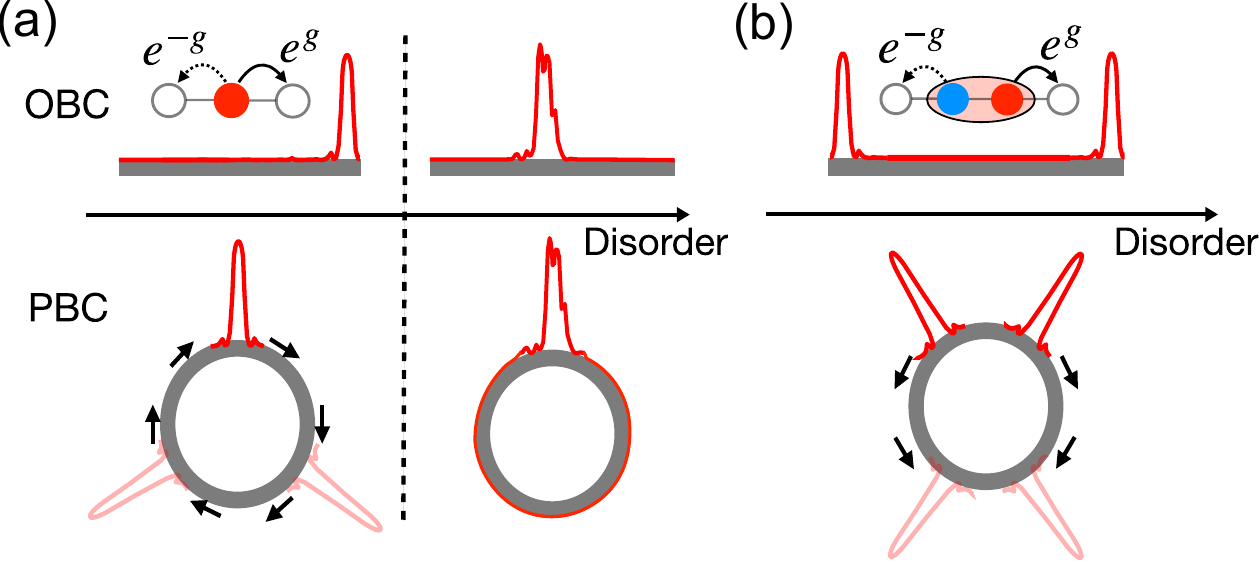}
\caption{(a) The charge skin effect (OBC) and persistent charge current (PBC) give way to Anderson localization as disorder strength is increased (b) The dipole skin effect (OBC) and persistent dipole currents (PBC) are stable to disorder. 
}
\end{figure}

Non-Hermitian systems host a variety of behaviors that are not possible in their Hermitian counterparts, ranging from generalized topological phases to novel entanglement transitions and criticality~\cite{Rudner_2009, Hu_2011, Borgnia2020, Lee2019Topological, Okuma2020Topological, Manna2023,Yao2018, Chen2020, Bergholtz2021,Tang_21,Li2018,*Li_Yao_19,Skinner19,Zabalo_2022,Chan2019,Schiro_2021}. One of the most celebrated phenomena is the non-Hermitian skin effect, where an extensive number of single-particle eigenstates are localized at the boundary of a lattice because of non-reciprocal hopping~\cite{Hatano1996,*Hatano1997,*Hatano1998NonHermitian,Brouwer1997,Feinberg1999}. 
Recently, the skin effect was generalized to systems that conserve multipole moments of a scalar U(1) charge~\cite{Gliozzi2024}.  Multipole conserving systems are intrinsically interacting, and the key signature of the \emph{multipole skin effect} in an $m$-pole conserving system is eigenstates with an extensive $(m+1)$-pole moment. For example, in dipole-conserving systems, the skin effect maximizes the quadrupole moment (see Fig~\ref{fig:1}(b)).

In their seminal works, Hatano and Nelson studied the effect of disorder on the non-Hermitian skin effect in the non-interacting case~\cite{Hatano1996, *Hatano1997, *Hatano1998NonHermitian}. They found that in open boundary conditions, the system undergoes a transition between two localized phases as a function of disorder strength. At weak disorder, the skin effect persists and wavefunctions are localized at a boundary; at strong disorder, the skin effect is overcome by Anderson localization, and wavefunctions are localized at random positions. When the boundary conditions are periodic, the non-reciprocal hopping creates a persistent current, and the skin effect phase becomes delocalized. In contrast, the Anderson-localized phase remains largely unaffected, since only a few localized modes are impacted by the change in boundary conditions (see Fig.~\ref{fig:1}(a)). Recently, the stability of this transition has been numerically investigated in interacting systems, which appear to have a non-Hermitian MBL phase~\cite{Hamazaki2019,Zhai2020Many,Suthar2022NonHermitian,Ghosh2022Spectral,DeTomasi2024Stable,Mak2024,Roccati2024}. {However, an analytical argument for this MBL phase is lacking, and the fate of intrinsically interacting systems, like those with multipole symmetry, remains unclear.}

Here we investigate the impact of disorder on systems that exhibit the non-Hermitian multipole skin effect. First, we generalize the Hatano-Nelson argument to the interacting case and show that, in the presence of U(1) charge conservation, there is a transition between a non-Hermitian skin effect phase and an MBL phase. In the dynamics, this corresponds to an area-to-volume law entanglement phase transition.
Next, we consider multipole-conserving systems and demonstrate that, for any amount of disorder, the multipole skin effect \emph{remains stable.} Remarkably, this implies that under periodic boundary conditions, all eigenstates are delocalized---even in the presence of strong disorder and stringent mobility constraints imposed by multipole conservation.

\textit{Charge Skin Effect}.---We begin by reviewing how disorder modifies the non-Hermitian skin effect for U(1) charge~\cite{Hatano1996, Brouwer1997, Feinberg1999} through the lens of the multipole skin effect~\cite{Gliozzi2024}. This approach clarifies the role of interactions and generalizes to more exotic skin effects for dipoles and higher multipoles.
The minimal model of non-reciprocal charge hopping in a disordered potential is the one-dimensional Hatano-Nelson Hamiltonian~\cite{Hatano1996},
\begin{equation}
	\label{eq:hn}
	H_\text{HN}(g) = \sum_{j=-L/2}^{L/2} (e^{g}c^\dag_{j+1}c_j + e^{-g}c^\dag_{j}c_{j+1}) + \sum_j h_j n_j,
\end{equation}
where $c^\dag_j$ is a fermion creation operator at site $j$, $h_j\in [-W, W]$ are random fields, $n_j=c^\dag_jc_j$, $g$ sets the strength of the non-Hermiticity, and $L$ is the system size. 

In the Hermitian limit $g=0$, all single-particle eigenstates are Anderson localized for any disorder~\cite{abrahams1979scaling}. 
In the clean limit ($W=0, g \neq 0$), the model exhibits the non-Hermitian skin effect: states localize at one edge in open boundary conditions (OBC) and delocalize in periodic boundary conditions (PBC)~\cite{Hatano1996}. Away from these limits, the interplay between disorder and non-reciprocity leads to a transition. For OBC, the transition separates a skin effect phase at weak disorder from Anderson localization at strong disorder, while for PBC it pits localization against delocalization.

\begin{figure}[t!]
\label{fig:2}
\hspace{-0.25cm}
\includegraphics[width=\linewidth]{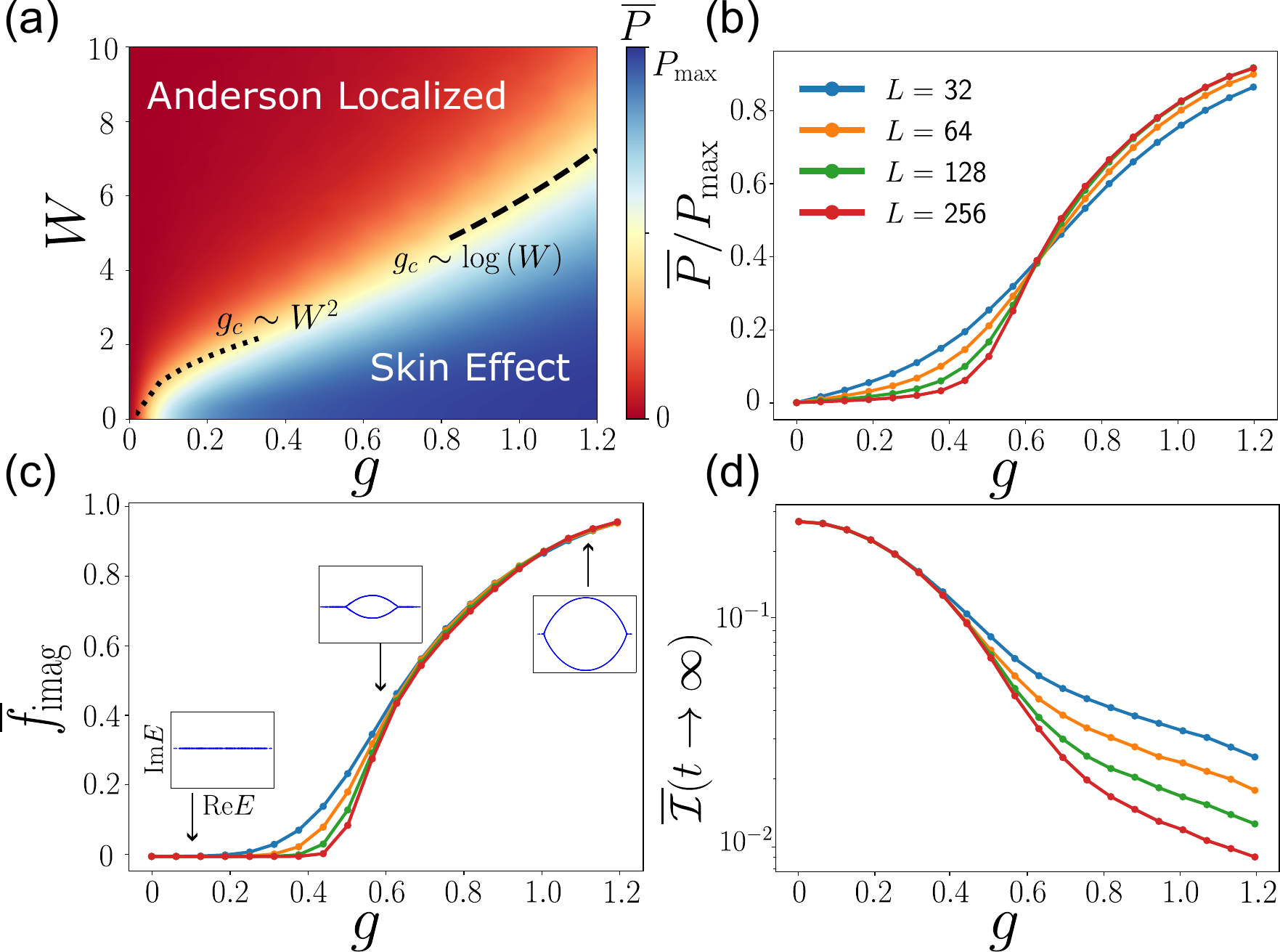}
\caption{\textbf{Non-interacting Hatano-Nelson model}, $W=4$. (a) Phase diagram as a function of disorder strength $W$ and non-Hermiticity $g$. {Dotted and dashed lines show the scaling of the transition at weak and strong disorder, respectively.} (b) An extensive dipole moment in many-body OBC eigenstates indicates the charge skin effect. (c) A finite fraction of PBC energies with a nonzero imaginary part,  $f_{imag}$, indicates delocalization (insets: sample single-particle spectra), (d) The imbalance $\mathcal{I}(t)$, after evolving a N\'eel state in PBC, does not go to zero in the localized phase at long times. Quantities with an overline are averaged over disorder and mid-spectrum eigenstates.
}
\end{figure}

This transition can be understood by making use of the similarity transformation~\cite{Hatano1996}
\begin{equation}
\label{eq:map}
	S^{-1} c^\dag_j S = e^{-gj} c^\dag_j, \quad S^{-1} c_j S = e^{gj} c_j,
\end{equation}
which maps the non-Hermitian model in OBC to its Hermitian version: $S^{-1} H_\text{HN}(g) S = H_\text{HN}(0)$. The existence of this transformation automatically implies that the OBC spectrum of $H_\text{HN}(g)$ is real. Furthermore, the right eigenstates of the non-Hermitian model can be obtained from the corresponding Hermitian ones: $\ket{\psi_g} = S\ket{\psi_0}$. 
Single-particle states are created by the operators
\begin{equation}
	\label{eq:orbital_h}
	\eta^\dag_\alpha (0) = \sum_j \phi_\alpha (j) c^\dag_j, \quad \eta^\dag_ \alpha(g) = S\eta^\dag_\alpha(0)  
 =  \sum_j e^{gj} \phi_\alpha(j) c^\dag_j,
\end{equation}
where the orbital $\phi_\alpha (j) \sim e^{-|j-j_\alpha|/\xi}$ is exponentially localized around a site $j_\alpha$ with localization length $\xi$~\footnote{In the non-Hermitian case, $\eta^\dag_\alpha(g)$ is only a creation operator for right eigenstates, such that $|{\psi_g^{(\alpha)}} \rangle = \eta^\dag_\alpha(g) |{0}\rangle$, and left eigenstates are created by a different operator, $\langle{\psi_g^{(\alpha)}}| = \langle{0}| \zeta_\alpha(g)$}. Thus, when $g < 1/\xi$, the non-Hermitian orbitals remain Anderson localized at an arbitrary distance from the boundary, while when $g>1/\xi$ they are localized on the right edge of the system and exhibit the skin effect. 
Figure~\ref{fig:2}(a) shows the phase diagram of this system as a function of $g$ and $W$. The localization length is energy-dependent, creating an intermediate region where localized and skin states are separated by a mobility edge. Moreover, at weak (strong) disorder, $\xi \sim 1/W^2$ ($\xi \sim 1/\log(W)$)~\cite{Derrida_84, Anderson1958}, leading to a critical non-reciprocity $g_c \sim W^2$ ($g_c \sim \log(W)$).

To establish an order parameter for the transition, we notice that the transformation in Eq.~\eqref{eq:map} is generated by
\begin{equation}
	\label{eq:map_dipop}
	S = e^{g \sum_j j n_j} = e^{g P},
\end{equation}
where $P$ is the total dipole moment operator.
At large $g$, $S$ tends to project onto states with the largest dipole moment~\cite{Gliozzi2024}. 
For single-particle states, this recovers the standard skin effect: clustering charge on the boundary maximizes the dipole moment. However, this map also naturally applies to many-body eigenstates, where the Pauli exclusion principle prevents multiple fermions from localizing at the boundary~\cite{Lee2020,Alsallom2022}. We thus define the U(1) charge skin effect more broadly as the tendency of eigenstates to maximize their dipole moment. In Fig.~\ref{fig:2}(b), the dipole moment renormalized by its maximum value serves as an order parameter for the OBC transition: in the Anderson phase it vanishes, whereas in the skin effect phase it does not.

We now turn to the case of periodic boundary conditions, in which the skin effect phase gives way to delocalization. Heuristically, if the barrier created by the edge is removed, the similarity transformation $S = e^{g P}$ pumps charge preferentially in one direction, leading to a persistent current $j \sim \partial_t P$. Without disorder, the energies of delocalized states form a loop in the complex plane: right-movers have a positive imaginary part and are amplified over time, while left-movers have a negative imaginary part and are suppressed.
On the other hand, at strong disorder, eigenstates are Anderson localized and therefore cannot be affected by boundary conditions. As a result, the energy of a localized eigenstate under PBC must be the same as its energy under OBC, and hence real. 
The delocalization transition is therefore associated with a real-to-complex transition of the PBC energy spectrum.
The fraction of eigenvalues with a non-zero imaginary part, shown in Fig.~\ref{fig:2}(c), is therefore a proxy for the transition in PBC. In the Anderson phase, $f_{\text{imag}} \rightarrow 0$, whereas in the delocalized phase, $f_{\text{imag}} \neq 0$. 

Finally, we can also identify the localization transition using the non-unitary dynamics defined by
\begin{equation}
\label{eq:psi_t}
\ket{\psi(t)} = \frac{e^{-i H_\text{HN}(g) t } \ket{\psi(0)}}{\lVert e^{-i H_\text{HN}(g) t } \ket{\psi(0)} \rVert}.
\end{equation}
Under PBC, time evolution is frozen in the Anderson localized phase, while a persistent unidirectional current carries charge in the skin effect phase. We contrast these behaviors by starting with a N\'eel state, $\ket{\psi(0)} = \prod_{i \in 2 \mathbb{Z}} c^\dag_i \ket{0}$, and tracking the imbalance of charge over time, $\mathcal{I}(t) = \sum_i (-1)^i \expval{n_i}{\psi(t)}$. In Fig.~\ref{fig:2}(d), we show the long time limit of the imbalance, which decays to zero for $g>g_c$ since the charges are mobile, but remains finite for $g<g_c$.

\textit{Interacting case}.---Having benchmarked the dipole moment as an order parameter for the transition, we now turn to the interacting case.
The interacting Hatano-Nelson model is defined as~\cite{Zhang2022,Kawabata2022,Hamazaki2019}
\begin{equation}\label{eq:hn_hubbard}
   H(g) = H_\text{HN}(g) + V \sum_j n_j n_{j+1}.
\end{equation}
As with any interaction that is diagonal in the occupation basis, the similarity transformation in Eq.~\eqref{eq:map} continues to map $H(g)$ to $H(0)$ in OBC. 

For sufficiently strong disorder, the interacting Hermitian system is believed to host an MBL phase~\cite{Oganesyan_2007,Pal_2010,Luitz_2015,Bera_2015,Devakul_2015,Sierant_2020,Vardhan_2017,GDT_2017,Doggen_2018}. The eigenstates in the MBL phase are adiabatically connected to those at $V=0$ by a quasi-local unitary transformation, and therefore maintain a similar spatial structure~\cite{Imbrie2016, ROS_2015, Huse_14, Chandran_2015}. However, these states cannot be expressed in terms of single-particle orbitals as in Eq.~\eqref{eq:orbital_h}, so it is convenient to work with conserved operators instead. 
In the non-interacting case, the local integrals of motion (LIOMs) associated with the right OBC eigenstates of Eq.~\eqref{eq:hn_hubbard} are given by
\begin{equation}
	\label{eq:liom_anderson}
\begin{aligned}
        \mathcal{O}^{(V=0)}_\alpha(0) := & \eta^\dag_\alpha(0) \eta_\alpha(0) = \sum_{i, j} \phi_\alpha(i) \phi_\alpha^*(j) c^\dag_i c_j \\ 
        \mathcal{O}^{(V=0)}_\alpha(g) := &\eta^\dag_\alpha(g) \eta_\alpha(g)= \sum_{i, j} e^{g (i+j)} \phi_\alpha(i) \phi_\alpha^*(j) c^\dag_i c_j.
\end{aligned}
\end{equation}
Since $\phi_\alpha(i) \phi_\alpha^*(j) \sim e^{-\text{max}(|i-j_\alpha|, |j-j_\alpha|)/\xi}$, the non-Hermitian LIOMs are randomly localized for $g < 1/\xi$ and edge localized for $g > 1/\xi$, indicating the Anderson and skin effect phases, respectively. 

\begin{figure}[t!]
\label{fig:3}
\includegraphics[width=\linewidth]{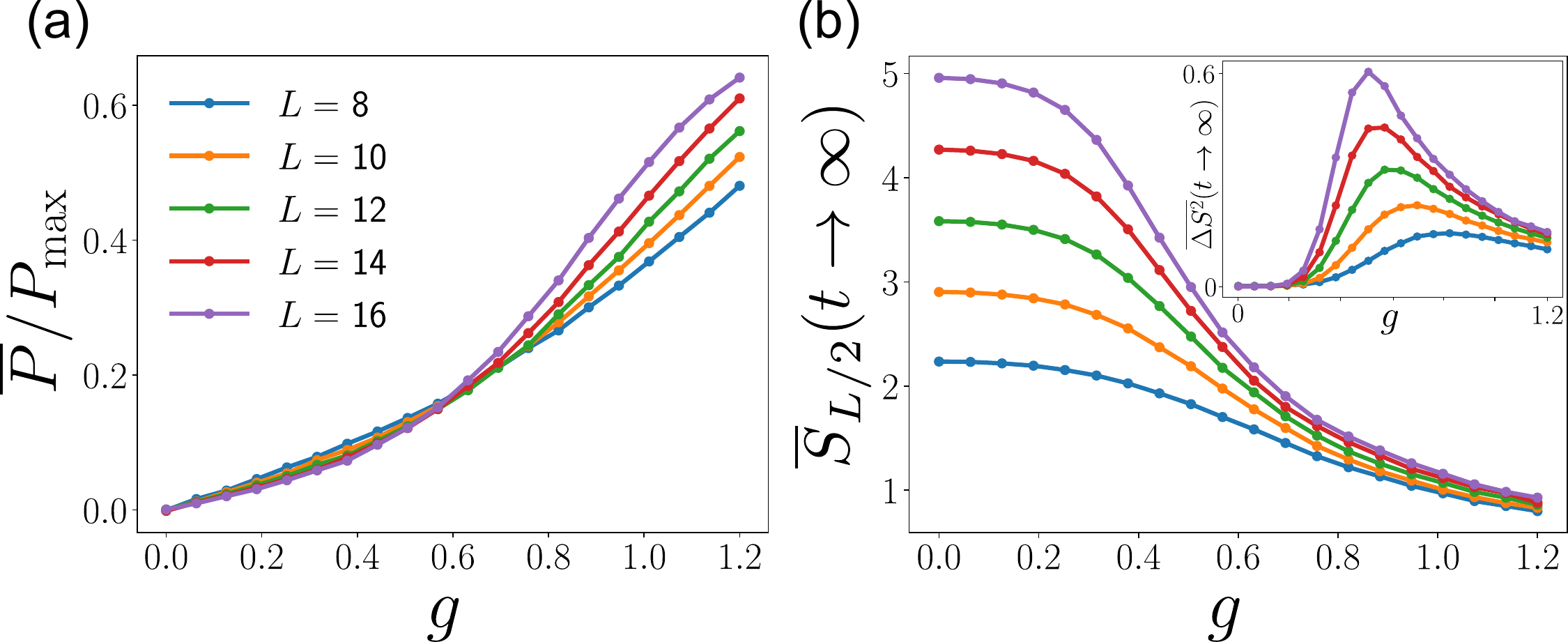}
\caption{\textbf{Interacting charge-conserving Hatano-Nelson model}, $V=1, W=8$ (a) Eigenstate dipole moment $P$ divided by its maximum value. For $g<g_c$ the system is in an MBL phase and $\overline{P}/P_{max}\rightarrow 0$, while this limit is finite in the skin effect phase for $g>g_c$. The dipole moment $P$ is averaged over disorder and mid-spectrum eigenstates. (b) Volume-to-area law entanglement transition in long-time steady states of the dynamics (inset: entanglement entropy fluctuations peak at the transition). The bipartite von Neumann entropy $S_{L/2}(t)$ is averaged over disorder and random initial states.
}
\end{figure}

In an MBL phase with $V>0$, LIOMs are locally dressed by the weak interactions:
\begin{equation}
	\label{eq:liom_mbl}
\begin{aligned}
	\mathcal{O}^{(V)}_\alpha(0) &= \sum_{i, j} \tilde{\phi}_\alpha(i, j)c^\dag_i c_j \\
	&+ \sum_{i, j, k, l} \tilde{\phi}_\alpha(i, j, k, l)c^\dag_i c^\dag_j c_k c_l + \ldots.
\end{aligned}
\end{equation}
These operators reduce to Eq.~\eqref{eq:liom_anderson} as $V\rightarrow 0$ and fall off exponentially away from the localization center $j_\alpha$, such that $\tilde{\phi}_\alpha(\{x_i\}) \sim e^{-\text{max}(|x_i - j_\alpha|)/\xi}$. 
All LIOMs commute with the total U(1) charge operator $N = \sum_j n_j$~\footnote{Since both the LIOMs and the charge commute with the Hamiltonian, if $[\mathcal{O}_\alpha, N] \neq 0$, then there would be a degenerate subspace of states, which is almost impossible due to the onsite disorder. See also Ref.~\onlinecite{Potter2016}.}.
Acting with the similarity transformation, we obtain the corresponding non-Hermitian (right-eigenstate) LIOMs:
\begin{equation}
	\label{eq:liom_mbl_g}
	\mathcal{O}^{(V)}_\alpha(g) = \sum_{n=1} \sum_{x_1, \ldots, x_{2n}} \!\! e^{g(x_1+\ldots + x_{2n})} \tilde{\phi}_\alpha(x_1, \ldots, x_{2n})c^\dag_{x_1} \ldots c_{x_{2n}}.
\end{equation}
In any given term of this expansion, the coefficient of the operator scales as $e^{g\max( x_i)} e^{-\text{max}(|x_i - j_\alpha|)/\xi}$, so localization occurs when $g < 1/\xi,$ and the skin effect occurs when $g > 1/\xi$. Thus, the many-body similarity transformation also explains the existence of non-Hermitian MBL, which was observed in Ref.~\onlinecite{Hamazaki2019}. 
We confirm the transition numerically {via exact diagonalization}, using the dipole moment as an order parameter (see Fig.~\ref{fig:3}(a)). When the system is periodic, the skin effect is replaced by a persistent current, and the transition is between MBL and delocalized eigenstates. This transition had been observed numerically in PBC~\cite{Hamazaki2019, Mak2024}, though its connection to the skin effect and its explanation in terms of LIOMs were not noted~\footnote{{Other works that numerically explore spectral and dynamical signatures of the non-Hermitian MBL transition include Refs.~\cite{Zhai2020Many, Suthar2022NonHermitian, Ghosh2022Spectral, DeTomasi2024Stable, Roccati2024}, though none discuss the physical mechanism for localization via non-Hermitian LIOMs.}}.
Note also that a version of the similarity transformation argument was previously used to extract the localization length in a Hermitian MBL system~\cite{Heussen2021,OBrien2023}.

Interestingly, although the transition connects two localized phases in an OBC system, interactions introduce a crucial difference in the dynamics of quantum entanglement~\footnote{We quantify entanglement via the half-partition von Neumann entanglement entropy,
$S_{L/2}(t) = -\text{Tr}[\rho_{L/2}(t) \log \rho_{L/2}(t)]$, with $\rho_{L/2}(t)$ the bipartite reduced density matrix of the evolved state $|\psi(t)\rangle$, starting from the N\'eel state.}. In the MBL phase, interaction-induced dephasing drives the system toward a volume-law-entangled state at long times. Conversely, in the skin-effect phase, the state is projected onto its component with maximal dipole moment, resulting in area-law entanglement~\cite{Gliozzi2024}. The MBL-to-skin-effect transition in OBC therefore induces an entanglement transition from volume- to area-law behavior, as shown in Fig.~\ref{fig:3}(b). {Because this entanglement transition relies on disorder, it is distinct from the one previously observed in clean non-Hermitian systems~\cite{Kawabata2023}.}

\textit{Multipole Skin Effect}.---We now turn to the skin effect in systems that conserve both a U(1) charge, like the total particle number $N = \sum_j n_j$, and its dipole moment $P = \sum_j j n_j$. 
A dipole-conserving generalization of the Hatano-Nelson model is 
\begin{equation}
\label{eq:h5}
H_5(g_0, g_1) = \!\!\!\! \sum_{r\in \{0,1\}, \, j} \!\!\!\! t_r (e^{g_r} c^\dag_j c_{j+1} c_{j+2+r} c^\dag_{j+3+r} + e^{-g_r} \times \mathrm{h.c.}),
\end{equation}
where the kinetic terms describe imbalanced dipole hopping as depicted in Fig.~\ref{fig:1}(b).
We consider the model at half-filling ($N=L/2$) and in the largest {sector of fixed dipole moment} ($P = 0$), and we initially concentrate on the clean limit. Without disorder, this system exhibits a \emph{dipole skin effect}, whereby dipoles are preferentially pumped to one boundary~\cite{Gliozzi2024}. Unlike the conventional (charge) skin effect, which clusters charges at one boundary, the dipole skin effect pushes charges symmetrically outwards to both edges of an open chain. In a system without boundaries, this leads to a persistent dipole current and delocalized states (schematically shown in Fig.~\ref{fig:1}(b)).

For the special case of Eq.~\eqref{eq:h5} with $g_1 = 3g_0/2$, the dipole skin effect can be understood via a similarity transformation analogous to Eq.~\eqref{eq:map_dipop}:
\begin{equation}
\label{eq:map_quadop}
S = e^{ g_0\sum_j j^2 n_j/4} = e^{g_0 \mathcal{Q}/4},
\end{equation}
where $\mathcal{Q}$ is the quadrupole moment of the charge distribution. This map transforms $H_5(g_0,3g_0/2) \mapsto H_5(0,0)$ via $(c^\dag_j, \, c_j) \mapsto (e^{-g_0j^2/4} c^\dag_j, \, e^{g_0j^2/4} c_j)$, enforcing a real spectrum in OBC.
Much like the charge skin effect produces eigenstates with an extensive dipole moment, the dipole skin effect leads to eigenstates with an extensive quadrupole moment. Moreover, these properties hold even when Eq.~\eqref{eq:map_quadop} does not map $H_5(g_0, g_1)$ to a Hermitian Hamiltonian~\cite{Gliozzi2024}.

Having discussed the skin effect in the clean limit, we return to systems with a disordered potential,
\begin{equation}
    \label{eq:hdip}
    H_\text{dip}(g) := H_5(g,3g/2) + \sum_j h_j n_j.
\end{equation}
At strong enough disorder~\footnote{For the chosen disorder strength $W=5.5$, we verify that the dynamics is frozen (see the inset of Fig.~\ref{fig:4}(c)) and that the level-spacing distribution is Poissonian. See Supplementary Information for additional plots}, the model is MBL when $g=0$~\cite{Herviou2021}.
In analogy with the charge-conserving case, we investigate the existence of a putative transition between MBL and the dipole skin effect at some $g_c\neq 0$. We provide evidence that this is not the case: any amount of non-Hermiticity spoils MBL and induces the dipole skin effect. In particular, this implies that the system with PBC is delocalized for \emph{any} disorder strength. 

\begin{figure}[t!]
\label{fig:4}
\includegraphics[width=\linewidth]{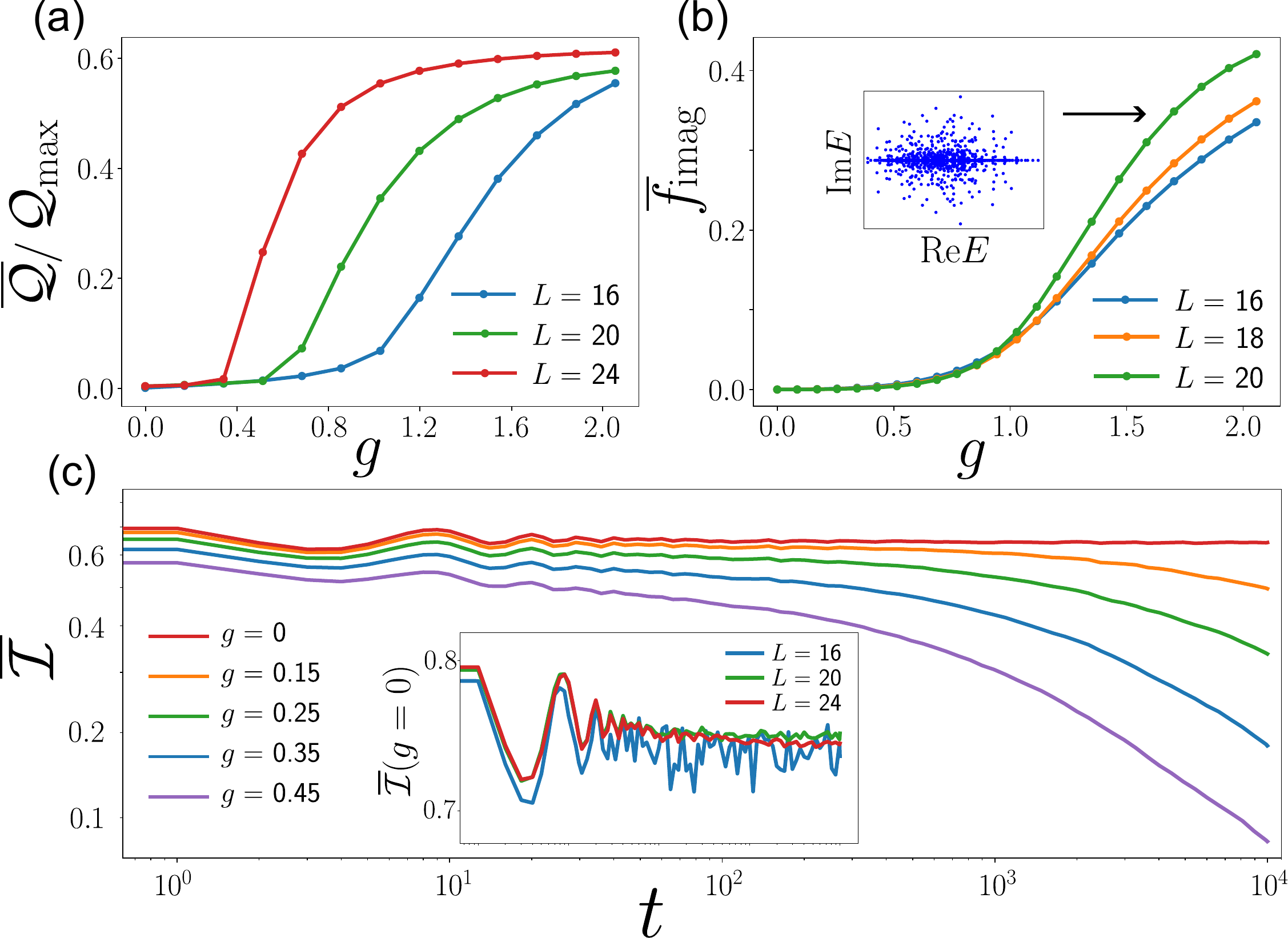}
\caption{\textbf{Dipole-conserving Hatano-Nelson model}, $W=5.5$, $t_1 = 0.3 t_0$. (a) The dipole skin effect manifests as extensive quadrupole moment in OBC eigenstates, and is present for any $g>0$ (b) The fraction of PBC energies with nonzero imaginary part tends to a finite value, indicating delocalization (inset: sample energy spectrum). Both quantities are averaged over disorder and mid-spectrum eigenstates. (c) Time evolution of imbalance starting from the N\'eel state in PBC with $L=24$. Imbalance remains finite at long times for $g=0$ (inset: different system sizes), while it decays for $g>0$.
}
\end{figure}

Our argument is very similar to the charge-conserving case (c.f.\ Eqs.~\eqref{eq:liom_mbl}--\eqref{eq:liom_mbl_g}) and relies on applying the similarity transformation to the Hermitian, dipole-conserving LIOMs. In the $g=0$ MBL phase, each state can be labeled by an extensive number of LIOMs,
\begin{equation}
    \mathcal{O}^\text{dip}_\alpha(0) = \sum_{i + k = j + l} {\varphi}_\alpha(i,j,k,l) c^\dag_{i} c_{j} c_k c^\dag_{l} + \ldots,
\end{equation}
where $\varphi_\alpha(\{x_i\}) \sim e^{-\text{max}(|x_i - x_\alpha|)/\xi}$ for some localization center $x_\alpha$, the dots indicate higher-order terms, and dipole symmetry is explicitly enforced in each term~\footnote{Diagonal terms like $\sum_i \varphi_\alpha(i) n_i$ are also included in the sum, as they conserve dipole moment}. Applying the map, we obtain the (right-eigenstate) LIOMs of the non-Hermitian system~\footnote{Right eigenstates define the non-Hermitian LIOMs, so creation and annihilation operators are scaled by the same factor, as in Eq.~\eqref{eq:liom_anderson}.}:
\begin{equation}
    \mathcal{O}^\text{dip}_\alpha(g) = \sum_{i+k=j+l} e^{g(i^2 + j^2 + k^2 + l^2)/4} {\varphi}_\alpha(i,j,k,l) c^\dag_{i} c_{j} c_k c^\dag_{l} + \ldots.
\end{equation}
Although $\varphi_\alpha$ is exponentially localized around a random lattice site $x_\alpha$, the prefactor $e^{g(i^2 + j^2 + k^2 + l^2)/4}$ is \emph{super}-exponentially localized at the outer edges of the chain. This latter factor always wins out, and for any $g>0$, the non-Hermitian LIOMs are pushed symmetrically to the boundaries of an OBC system. Consequently, the dipole skin effect always prevails over MBL. 

Heuristically, there is a competition between two tendencies: localizing charges randomly and localizing dipoles on the boundary. However, charges are zero-dimensional objects while dipoles are one-dimensional. The dipole skin effect can thus arise both from fixed-length dipoles moving to the edge and from the boundary dipoles themselves growing in length. 
The latter process is unique to dipoles, and provides extra pathways to maximize the quadrupole moment. It is this additional localization mechanism that allows the dipole skin effect to dominate over the localization of charge~\footnote{See Supplemental Material for a case in which dipoles are local, zero-dimensional objects, leading to a transition}.

We verify this numerically by plotting the average quadrupole moment of the eigenstates of $H_\text{dip}(g)$ in Fig.~\ref{fig:4}(a): the average quadrupole moment grows with system size for any $g\ne0$. In PBC, the skin effect becomes a persistent dipole current that delocalizes the system. To see this, we plot the fraction of complex energies $f_{\text{imag}}$ in Fig.~\ref{fig:4}(b), which tends to a finite value with increasing system size. We also investigate the out-of-equilibrium dynamics through the imbalance $\mathcal{I}(t)$ starting from the previously defined N\'eel state, shown in Fig.~\ref{fig:4}(c). After a transient, the dynamics is frozen for $g=0,$ but mobile for any amount of non-Hermiticity. 

The primacy of the dipole skin effect over MBL extends to the broader family of multipole skin effects~\cite{Gliozzi2024}. 
The $m$-pole skin effect involves $m$-poles preferentially hopping to one side to maximize the $(m+1)$-pole moment. Repeating our previous arguments, the similarity transformation that achieves this is $S \sim e^{g \sum_j j^{m+1} n_j}$. Hermitian LIOMs, which are at most exponentially localized, are pushed away from their localization centers by the non-Hermiticity. 
If $m\geq 1$, the multipole skin effect overcomes disorder in OBC, and delocalization ensues in PBC. Beyond multipole conservation, the skin effect can be generalized for any modulated symmetry charge $\mathcal{Q} = \sum_j f(j)\, n_j$~\cite{Sala2022}, where the scaling of $\mathcal{Q}$ determines its localization behavior.

\textit{Discussion}.---We have seen that the fate of localization in disordered, non-Hermitian systems strongly depends on the underlying symmetries.
Non-reciprocal systems that conserve a U(1) charge host a transition between localization and the non-Hermitian skin effect. 
Under PBC, the skin effect becomes a persistent current and this is a genuine delocalization transition.
On the other hand, when a multipole moment of the charge is also conserved, the skin effect always dominates over localization. 
Consequently, a PBC system with multipole conservation, which imposes kinetic constraints that hinder particle motion~\cite{Sala2020, Feldmeier2020, Moudgalya2020}, and strong disorder, which also tends to freeze particles, is nevertheless delocalized. Surprisingly, non-hermiticity is enough to circumvent both disorder and kinetic constraints and evade localization.

One consequence of our results is that multipole skin effects can be stabilized even when multipole conservation is explicitly broken by charge hopping: there exists a regime where disorder localizes the charge while multipoles remain delocalized. {This stability is important for experimental realizations because dipole-conservation is typically only an approximate symmetry. For example, optical lattices with an external tilt effectively conserve dipole moment for long timescales~\cite{Guardado-Sanchez2020, Scherg2021, Kohlert2023}. Onsite disorder is easily implemented in such cold atom systems, and the addition of external loss can induce non-reciprocal couplings~\cite{Gong2018}, including multipole-hopping~\cite{Hu2025Dec}. Metamaterials, such as topoelectric circuits~\cite{Lee2018} and acoustic lattices~\cite{Jing2024}, provide another candidate platform, and allow for strong non-reciprocities of up to $g \sim 1.5$.}

Relatedly, it would be interesting to investigate possible transitions between skin effects for different multipoles, like dipoles and charges. The effects of discrete and non-Abelian symmetries on the disordered skin effect should also be explored. Finally, moving beyond one-dimensional systems remains an important avenue for further study.
\\~\\
\noindent
\textit{Acknowledgements}.---J.G. and T.L.H. acknowledge support from the US Office of Naval Research MURI grant N00014-20-1-2325.
G.D.T. acknowledges support from the EPiQS Program of the Gordon and Betty Moore Foundation and
the hospitality of MPIPKS Dresden, where part of the work was done. 

\bibliographystyle{apsrev4-1}
\bibliography{main.bib}

\setcounter{figure}{0}
\renewcommand{\thefigure}{S\arabic{figure}}
\setcounter{equation}{0}
\renewcommand{\theequation}{S\arabic{equation}}

\clearpage
\begin{center}
    \large\textbf{Supplementary Material}
\end{center}

\noindent
{\bf SM 1: Skin effects without the similarity transformation to a Hermitian Hamiltonian}
\\
\indent
In the main text, we focused on non-Hermitian systems with a real spectrum in open boundary conditions. In these cases, a similarity transformation maps the non-Hermitian Hamiltonian to a Hermitian one. For both the charge skin effect and the dipole skin effect, our arguments about stability to disorder were based on applying this similarity transformation to obtain non-Hermitian eigenstates from their Hermitian counterparts.

Nevertheless, it was argued in Ref.~\onlinecite{Gliozzi2024} that even Hamiltonians with a complex OBC spectrum exhibit the same skin effect phenomenology. Here, we show that this is also true for these Hamiltonians with onsite disorder. Concretely, we consider non-reciprocal models that do not admit a similarity transformation to a Hermitian system, and numerically show that they have the same key features and localization transitions as the models in the main text.

\begin{figure}[t!]
\label{fig:S1}
\includegraphics[width=\linewidth]{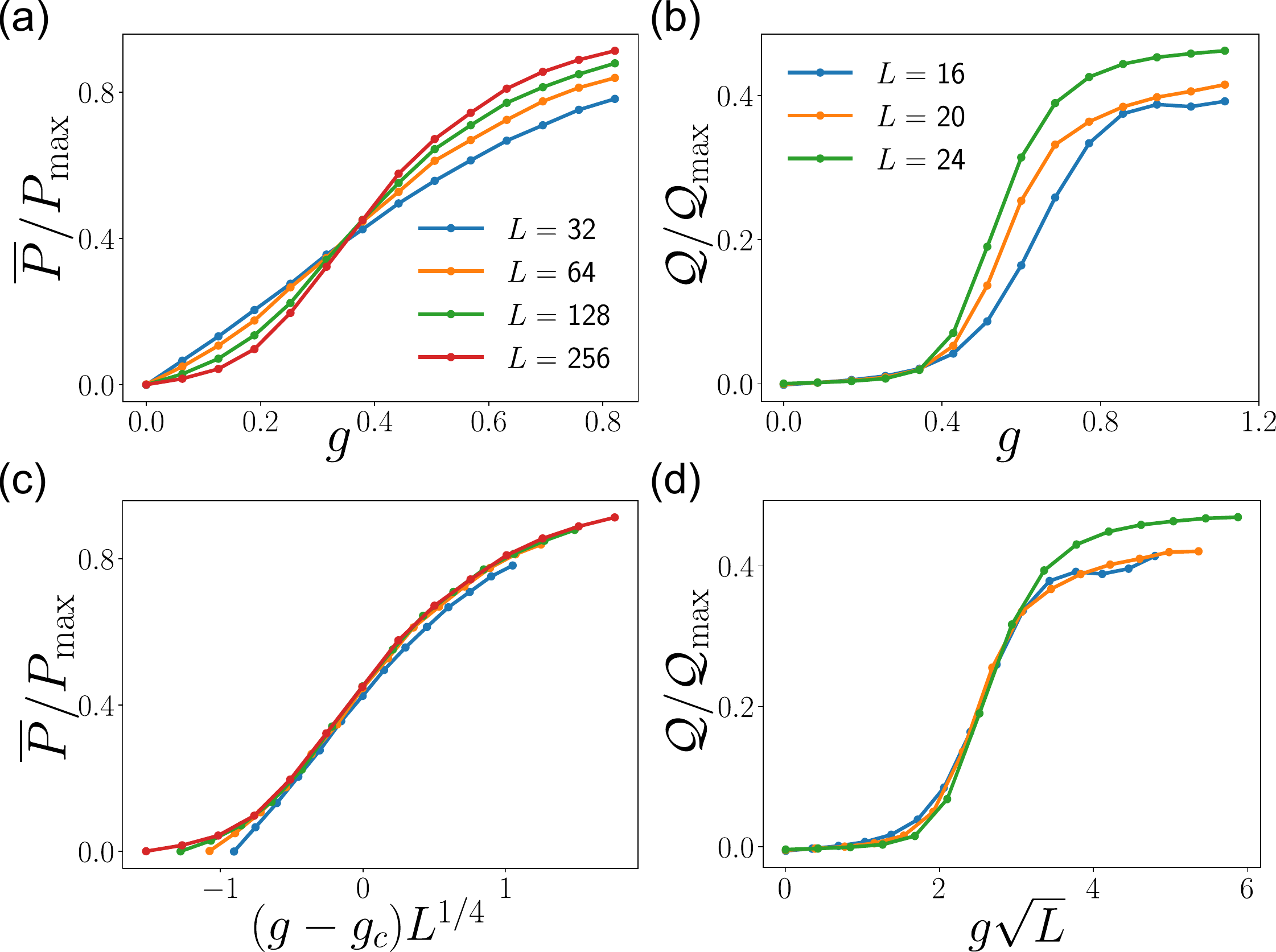}
\caption{Skin effect order parameters for models without a similarity transformation (a) The charge-conserving Hatano-Nelson model with next-nearest neighbor hoppings, $\tilde{H}_\text{charge}(g)$, has a transition between Anderson localization and the skin effect as measured by eigenstate dipole moment ($W=4$). (b) The dipole-conserving model $\tilde{H}_\text{dip}(g)$ appears to exhibit the dipole skin effect for any $g>0$, as measured by eigenstate quadrupole moment. The data collapse reasonably well when non-reciprocity is rescaled by (c) $L^{1/4}$ ($g_c = 0.38$) for the charge-conserving case and (d) $\sqrt{L}$ ($g_c = 0$) for the dipole-conserving case.} 
\end{figure}

For U(1) charge conservation, we can ``break'' the similarity transformation by adding a next-nearest neighbor hopping term to the Hatano-Nelson model:
\begin{equation}
    \label{eq:ham_nomap_charge}
    \tilde{H}_\text{charge} = -\sum_j [e^g (c^\dag_{j+1} c_{j} + c^\dag_{j+2} c_j) + e^{-g} \times \mathrm{h.c.}] + \sum_j h_j n_j,
\end{equation}
where again the disordered potential $h_j \in [-W, W]$. Even in open boundary conditions, this Hamiltonian does not have a real spectrum for $g>0$, and therefore does not admit a similarity transformation to a Hermitian model. Nevertheless, we numerically verify in Fig.~\ref{fig:S1}(a) that it still has a transition between an Anderson localized phase and a phase exhibiting the charge skin effect. In Fig.~\ref{fig:S1}(d), show an approximate data collapse when the non-reciprocity is rescaled by $L^{1/4}$.

In the dipole-conserving case, we can ``break'' the similarity transformation by choosing different imbalances $g_0$ and $g_1$ for range-four and range-five dipole hopping terms:
\begin{equation}
    \label{eq:ham_nomap_dip}
    \tilde{H}_\text{dip}(g) = H_5(g, 2g) + \sum_j h_j n_j,
\end{equation}
This Hamiltonian has a complex spectrum in open boundary conditions and does not admit a similarity transformation to a Hermitian model. Its eigenstates cannot therefore be obtained by applying $S = e^{g \mathcal{Q}/4}$ to a set of Hermitian eigenstates. However, we numerically verify in Fig.~\ref{fig:S1}(b) that it still exhibits the dipole skin effect for any $g>0$, as in the case where the map works. In Fig.~\ref{fig:S1}(d), we also notice an approximate data collapse when the non-reciprocity is rescaled by $\sqrt{L}$.

In general, our predictions for localization transitions or lack thereof should also be valid for systems in which the similarity transformation does not work, as long the direction of non-reciprocity remains the same. Fundamentally, all multipole skin effects are driven by an imbalanced hopping of $m$-poles. Independently of whether a similarity transformation exists, such hopping creates a current of $m$-poles that tends to maximize the $(m+1)$-pole moment in an open system. 

\noindent
{\bf SM 2: Additional evidence for many-body localization in the Hermitian dipole-conserving model}
\\
\indent
In the main text, we considered a dipole-conserving model with onsite disorder $W=5.5$, such that the Hermitian version, $H_\text{dip}(g=0)$, is many-body localized. We showed dynamical evidence for MBL in Fig.~4(c), where the long-time imbalance starting from the Néel state saturates at a nonzero value when $g=0$. Here we present further spectral evidence that the Hermitian system with $W=5.5$ is indeed many-body localized, focusing on the symmetry sector containing the N\'eel state.

\begin{figure}[t!]
\label{fig:S2}
\includegraphics[width=\linewidth]{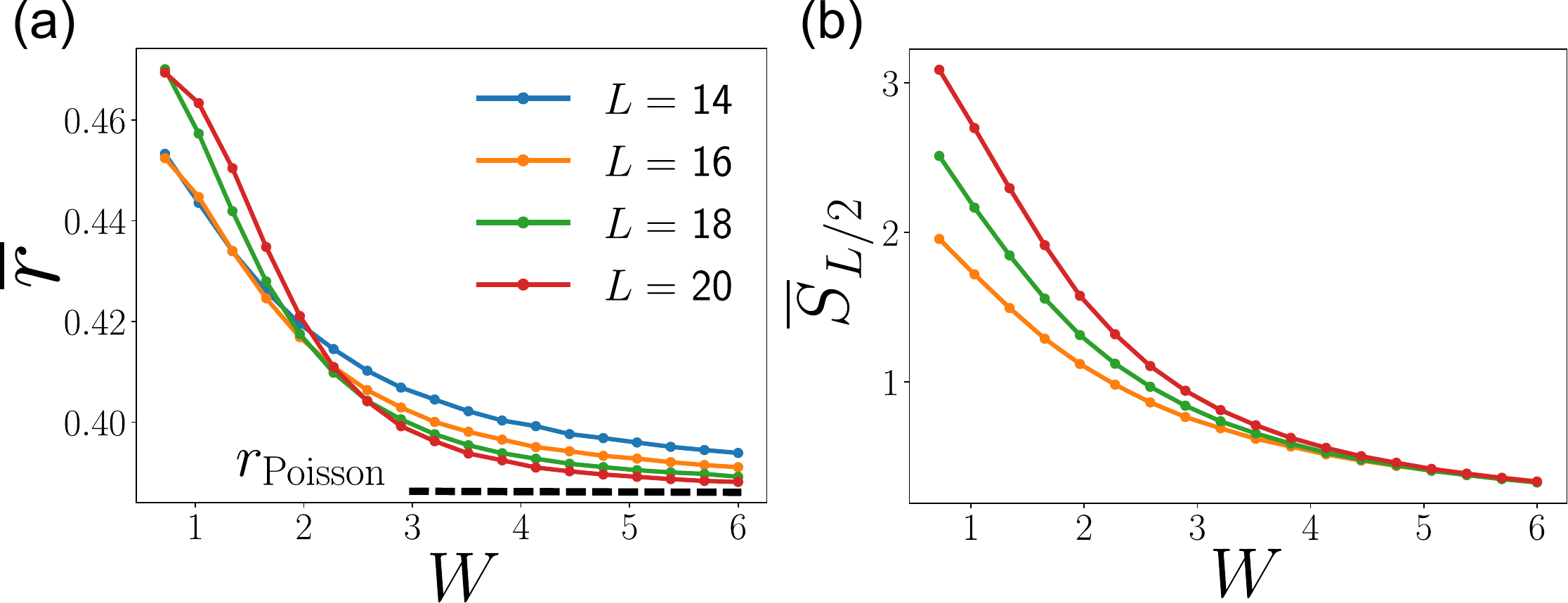}
\caption{Spectral evidence of many-body localization in the Hermitian, dipole conserving model $H_\text{dip}(0)$. (a) The spectral statistics of midspectrum energy levels are Poissonian at $W=5.5$, (b) The half-chain entanglement entropy of midspectrum eigenstates is area law scaling at $W=5.5$.}
\end{figure}

The first signature we consider is the energy level statistics. Chaotic Hamiltonians exhibit level repulsion, and midspectrum energies are distributed according to the Gaussian Orthogonal Ensemble (GOE). On the other hand, MBL systems do not have level repulsion, and midspectrum energies have a Poissonian distribution. The proximity to either limit can be quantified by ratio of consecutive energy gaps,
\begin{equation}
    r_n = \frac{\min (s_n, s_{n+1})}{\max (s_n, s_{n+1})}, \quad s_n = E_{n+1}-E_n.
\end{equation}
The average value of $r_n$ (over disorder and eigenstates) is called the r-statistic $\overline{r}$. In MBL systems, we expect a value close to $r_\text{Poisson} \approx 0.386$, while in a chaotic system we expect $r_\text{GOE} \approx 0.531$. In Fig.~\ref{fig:S2}(a), the r-statistic at $W=5.5$ is very close to Poissonian, indicating an MBL phase.

The second signature we consider is the entanglement entropy of mid-spectrum eigenstates. Concretely, we consider the von Neumann entanglement entropy for a bipartition of our system, $S_{L/2} = -\text{Tr}[\rho_{L/2} \log \rho_{L/2}]$, where $\rho_{L/2}$ is the reduced density matrix of an eigenstate on half of the chain.
In a chaotic system, these states are expected to be thermal and therefore have a volume-law entanglement, $S_{L/2} \sim L$. In an MBL system, midspectrum eigenstates are non-thermal and area-law entangled, $S_{L/2}\sim \mathcal{O}(1)$. In Fig.~\ref{fig:S2}(b), the entanglement entropy at $W=5.5$ appears to not scale with system size, indicating area-law entanglement.

Lastly, we confirm that the Hermitian MBL phase at $W=5.5$ has a set of exponentially localized conserved operators. These operators correspond to the LIOMs of Eq.~(8) in the main text, and their localization in the Hermitian phase is a key property of MBL. We construct these LIOMS numerically by following Ref.~\cite{Chandran_2015} and taking the long-time average of a local operator:
\begin{equation}\label{eq:time_avg}
    \tilde{n}_i := \lim_{T\rightarrow \infty} \frac{1}{T} \int_0^T n_i(t),
\end{equation}
where $n_i(t)$ denotes the Heisenberg evolution of the local density under the Hamiltonian. After time averaging, only terms diagonal in the energy basis survive:
\begin{equation}\label{eq:numerical_liom}
    \tilde{n}_i = \sum_{E} \expval{n_i}{E} \ket{E} \bra{E}.
\end{equation}
Here the sum is taken over the entire eigenstate basis of $H_\text{dip}(0)$, and $\tilde{n}_i$ is a conserved operator by construction. Furthermore, the support of $\tilde{n}_i$ is concentrated near site $i$ in the MBL phase. The spread of this LIOM can be quantified via the two point correlation function,
\begin{equation}\label{eq:Mij}
    M_{ij} = \frac{1}{\mathcal{D}} \Tr(\tilde{n}_i n_j) = \frac{1}{\mathcal{D}} \sum_{E} \expval{n_i}{E} \expval{n_j}{E},
\end{equation}
where $\mathcal{D} = 2^L$ is the Hilbert space dimension. This correlator can be interpreted as the effect that a charge at site $i$ has on site $j$ at late times. In an MBL phase, this effect falls off exponentially as $M_{ij} \sim e^{-|i-j|/\xi}$, coinciding with the exponential localization of the LIOM $\tilde{n}_i$.

\begin{figure}[t!]
\label{fig:S3}
\includegraphics[width=\linewidth]{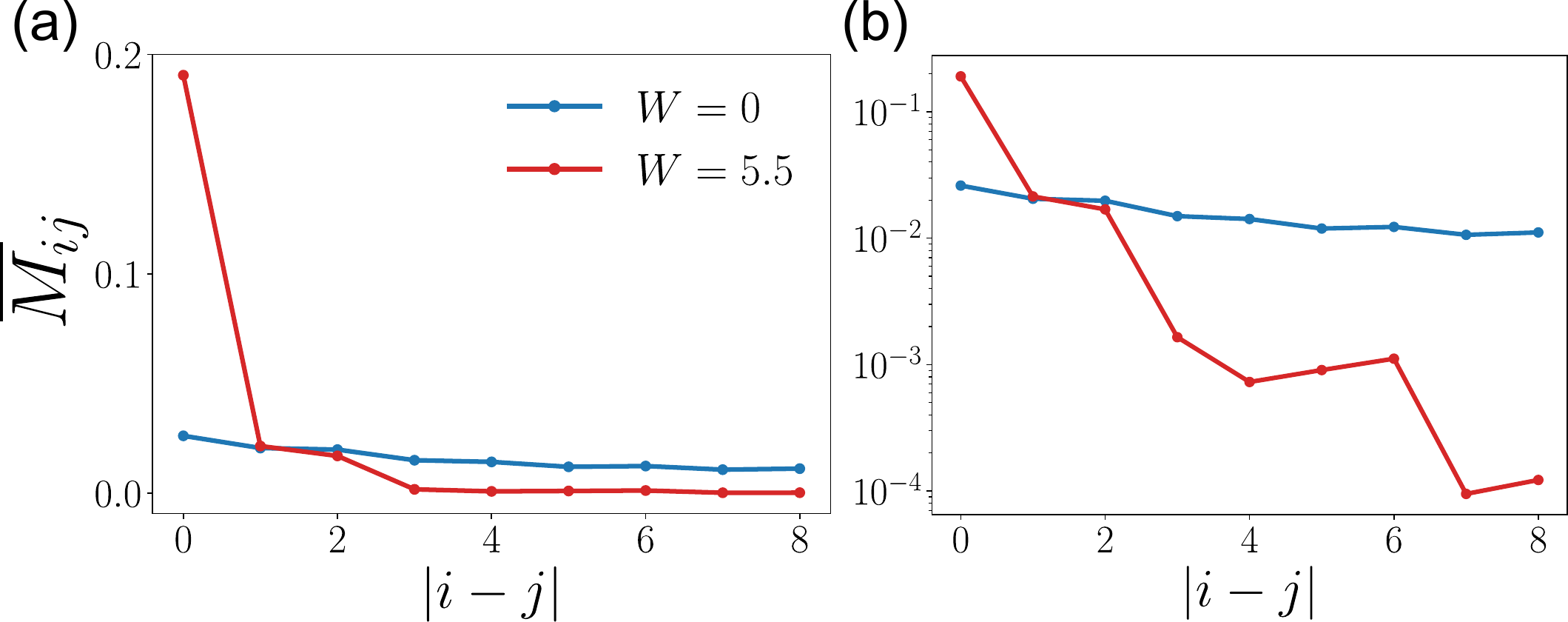}
\caption{Spatial profile of conserved operators $\tilde{n}_i$ of $H_\text{dip}(0)$ in a periodic chain with $L=18$. In the clean system (blue), these operators are extended, while in the MBL phase at $W=5.5$ (red) they are exponentially localized LIOMs. The results are shown with both (a) linear and (b) logarithmic scales, and are averaged over disorder and lattice sites.}
\end{figure}

We show the behavior of $M_{ij}$ in both the MBL phase and the clean limit in Fig.~\ref{fig:S3}. The results are averaged over 32 disorder realizations and over different sites in a periodic chain of length $L=18$. For a LIOM defined using the full energy eigenbasis as in Eq.~\eqref{eq:numerical_liom}, $M_{ij} \rightarrow 0$ at large $|i-j|$~\cite{Chandran_2015}. However, we construct the $\tilde{n}_i$ by only summing over the eigenstates in the symmetry sector of the N\'eel state.
In other words, we actually calculate 
\begin{equation}\label{eq:actual_Mij}
M'_{ij} = \frac{1}{\mathcal{D}_\text{sector}} \sum_{E \in \text{sector}} \expval{n_i}{E} \expval{n_j}{E}.
\end{equation}
Because of the symmetry constraints, $M'_{ij}$ does not go to zero at large distances, and we must first subtract a background to see the exponential decay. This background is given by
\begin{equation}\label{background}
B = \lim_{|i-j|\rightarrow \infty} \frac{1}{\mathcal{D}_\text{sector}} \Tr_\text{sector}(n_i n_j),
\end{equation}
the long-distance correlator between two operators that are strictly local in the full Hilbert space, but slightly correlated within the symmetry sector. We therefore plot $M_{ij} = M'_{ij} - B$, which shows an exponentially localized LIOM in the localized phase. In the clean phase, on the other hand, the putative LIOM $\tilde{n}_i$ is no longer local.

\noindent
{\bf SM 3: Localization to skin effect transition in a model with hidden U(1) symmetry}
\\
\indent
We have argued that non-reciprocal U(1) charge hopping permits a transition between disorder-induced localization and the skin effect. On the other hand, non-reciprocal dipole hopping, which conserves both U(1) charge and its dipole moment, always leads to the dipole skin effect and never many-body localization. 

Here we discuss a special case that naively appears to forbid a localization transition but actually hosts one because of a hidden U(1) symmetry. Consider a dipole-conserving Hamiltonian with only the minimal-range hopping term:
\begin{equation}
\label{eq:h4}
H_4(g) = \sum_j (e^g c^\dag_j c_{j+1} c_{j+2} c^\dag_{j+3} + e^{-g} \times \mathrm{h.c.}) + \sum_j h_j n_j.
\end{equation}
Let us restrict our attention to the symmetry sector at half-filling $N=L/2$ and dipole moment $P=0$. Inside this sector, the limited range of the dipole hopping imposes a strong kinetic constraint: many states are not connected to each other by the Hamiltonian. In fact, the sector itself fragments into exponentially many disconnected Krylov subsectors, and the largest of these is exponentially small in system size ($D_\text{largest}/D_\text{total} \sim e^{-L}$)~\cite{Sala2020, Moudgalya2020}. While in the main text we avoided this (strong) \emph{Hilbert space fragmentation} by adding a longer-ranged dipole hopping term, it is also interesting to consider how the skin effect and disorder are manifest in individual Krylov subsectors.

The largest Krylov subsector is integrable and can be mapped to a free-fermion chain of length $L/2$~\cite{Moudgalya2020}. The mapping involves grouping sites $2k-1$ and $2k$ of the original lattice into a new composite site. After this rewriting, the integrable subsector is the one where each pair of sites contains only a single occupied state. In other words, only the computational basis states $\ket{\rightarrow}:=\ket{01}$ and $\ket{\leftarrow}:=\ket{10}$ are allowed on every (odd-even) pair of sites. These two states represent a local dipole and a local anti-dipole, respectively.
For a composite site $k:=(2k-1, 2k)$, we can define dipole creation and annihilation operators, $d^\dag_k := c_{2k-1} c^\dag_{2k}$ and $d_k := c^\dag_{2k-1} c_{2k}$. These operators transform a local dipole into a local anti-dipole and vice versa. When acting on the integrable subspace, they obey fermionic anticommutation relations, so they create genuine local quasiparticle excitations. Furthemore, if we define a number operator for these local dipoles, $n^d_k = d^\dag_k d_k$, then the total dipole operator in the subsector can be written as
\begin{equation}\label{eq:local_dip}
    P = \sum_{j=1}^{L} j n_j = \sum_{k=1}^{L/2} n^d_k, 
\end{equation}
where the first expression is on the original lattice and the second expression uses the composite degrees of freedom. In this sector, the conservation of dipole moment acts as an ordinary U(1) symmetry with local charges.

Since only the odd terms of \eqref{eq:h4} act nontrivially on the subspace, we can project the Hamiltonian into the subsector using the composite degrees of freedom:
\begin{equation}\label{eq:h4_sector}
\begin{aligned}
 H^\text{sub}_4(g) = \sum_k (e^g d^\dag_{k+1} d_{k} + e^{-g} d^\dag_{k} d_{k+1}) + \sum_k \tilde{h}_k n^d_k, 
 \end{aligned}
\end{equation}
where $\tilde{h}_k := h_{2k} - h_{2k-1}$ is an onsite potential derived from the action of the original disorder on the new degrees of freedom, and we have ignored an overall constant.
This Hamiltonian is identical to the Hatano-Nelson model, though with the interpretation that the charges are actually local dipoles. Moreover, $H^\text{sub}_4(g)$ maps to its Hermitian counterpart under the similarity transformation generated by $S = e^{g\sum_k k n^d_k}$, which maximizes the dipole moment of the local dipoles. As a result, it hosts a transition between Anderson localization and the ``charge skin effect'', where both types of localization involve local dipole quasiparticles. 

Such a transition can only occur if the dipoles are local quasiparticles as in this Krylov subsector. Generally, dipoles are 1D objects and their localization (via the dipole skin effect) dominates over the localization of 0D charge in a possible Anderson/MBL phase. However, if dipoles can be mapped to local charges, then, like charged particles, their Anderson localization competes with the skin effect. This is a highly fine-tuned situation. For the dipole moment to decompose into a sum of local densities as in Eq.~\eqref{eq:local_dip}, the Hilbert space must be truncated to forbid states containing longer dipoles, and the Hamiltonian must be carefully tuned to preserve these states. For example, even $H_4(g)$ does not exhibit this behavior in general, and only a vanishingly small subsector of the Hilbert space hosts a transition.

The hidden U(1) symmetry manifests in another way: the dipole skin effect is more weakly localized than usual inside the Krylov subsector.
The dipole skin effect maximizes the quadrupole moment, which generically scales as $L^3$ for many-body states, leading to the super-exponential localization of LIOMs discussed in the main text. In this restricted sector, however, any quadrupole must be formed by local dipoles, and the largest quadrupole moment, associated with the state $\ket{\leftarrow \leftarrow \ldots \rightarrow \rightarrow} = \ket{1010\ldots 0101}$, only scales as $L^2$. This weaker scaling implies only exponential localization of LIOMs from the skin effect, and thus allows a localization transition.

\end{document}